\documentstyle[12pt,epsfig]{article}

\newcommand{\win}{\displaystyle W_{in}}
\newcommand{\wou}{\displaystyle W_{out}}
\newcommand{\tso}{\displaystyle T_{av}}
\newcommand{\tpr}{{W}/{m}\cdot K}
\newcommand{\tgr}{{K}/{mm}}
\newcommand{\gr}{\displaystyle \left|\nabla T\right|}
\begin{document}
\large

\begin{center}
{\Large \bf 
ESR Spectrometry of High-Temperature Superconductors 
with Temperature-Modulation-Based Thermoregulation}
\end{center}

\begin{center}
{\bf 
M.K. Aliev, G.R. Alimov\footnote[1]{Corresponding author. 
e-mail: gleb@iaph.silk.org }, I.Kholbaev,\\
T.M. Muminov, B. Olimov, B.Yu. Sokolov,\\
and R.R. Usmanov,}
\end{center}

\begin{center}
\small Institute of Applied Physics, Tashkent State University
700095, Tashkent, Uzbekistan \\
\end{center}

\begin{abstract}
A light-beam-assisted temperature-control system operating within a 
temperature range from 77 to 180 K is developed for investigating 
the absorption of electromagnetic waves by high-temperature 
superconductors in the vicinity of superconducting transition with 
an ESR spectrometer. The advantage of this system is the feasibility 
of modulating the temperature of a sample with a frequency of 80 Hz
and amplitude of $ 10^{-2}\div 10^{-1}K $. The rms temperature 
instability over a 
5-min time interval is within 0.06 K, the temperature gradient in
a sample is $ \sim 10^{-2} K/mm $  for $ T \sim 90K $, 
and the system relaxation time 
is 1 -- 10 s. The microwave absorption peaks in the vicinity of the 
superconducting transition of an $YBa_2Cu_3O_{7-x}$, 
single crystal were first 
measured at (the simultaneous application of magnetic and temperature 
modulations.
\end{abstract}

\vspace{1mm}

\section{Introduction}
\indent

It is known that the temperature dependence of
microwave response in high-temperature superconductors (HTSCs), 
which is recorded by ESR spectrometers
using the magnetic modulation technique, peaks at the
superconducting transition temperature \cite{ref1} - \cite{ref4}. 
This signal can serve as a source of information on the features
of the superconducting transition in HTSCs. The observation 
of this signal, when temperature modulation is
used, and its comparison with the signal recorded with
the use of magnetic modulation are of significant interest. 
Note that ESR spectrometers with flow-type cryostats \cite{ref5} do not 
provide the necessary temperature uniformity over a sample \cite{ref6} 
and, in addition, it is difficult
to modulate the temperature of a sample in them.

Below we describe a temperature-control system
designed to study HTSCs at an SE/X 2543 RADIOPAN
ESR spectrometer. The sample temperature is controlled
with a light beam. This system has significantly better
characteristics in a temperature range of $ 77 < T < 120 K $
than flow-type cryostats, and, importantly, allows us to
modulate the sample temperature with a required frequency.

\section{Method}
\indent
 
The left part of Fig. 1 shows the general view of our
cryosystem. Figure I (central part) shows the design of
the sample holder (SH), which is one of the most
important elements that determine technical parameters
of the cryosystem.

The working volume of the holder is represented by
a thin-walled capsule 1. It is placed inside a fingerlike
extension of a quartz dewar (D) with liquid nitrogen,
which is inserted into the microwave resonator (MR)
through an opening in its top wall. The SH dimensions
are determined mainly by those of the resonator and
dewar D.

Sample 2 and the working junction of a differential
chromel-copel thermocouple 3 (wires have a diameter
of 0.15 mm) being inside the capsule are placed inside
heat exchanger 4. The latter is a sapphire cylinder
sawed into two parts, as shown in Fig. 1. The main,
larger part of the heat exchanger has depressions, into
which the working junction of the thermocouple with
wires is pressed. To ensure a better thermal contact with
the heat exchanger, hollow cavities are filled with a
mixture of a finely dispersed corundum powder with a
silicate adhesive.

The reference junction of the thermocouple is
placed in a separate dewar with liquid nitrogen (not
shown in Fig. 1), where its temperature is monitored to
an accuracy of $\pm 0.005K$ with a 650H UNIPAN platinum 
resistance thermometer. The chromel-copel thermocouple 
has a sensitivity of $\ge 30 \mu /K$ at the liquidnitrogen 
temperatures. 
This makes it possible to monitor the temperature variation for 
the working junction
of the thermocouple to an accuracy of $\pm 0.03 K$.

The main part of the heat exchanger has a saw cut
0.3 mm wide and 1 mm deep along the cylinder diameter, in which 
the sample under study is mounted (as a
rule, samples of HTSC single crystals have the form of
plates with a width of a few tens of microns). The sample is 
separated from the working junction of the thermocouple in order 
to avoid the influence of local distortions of the microwave 
field caused by the junction. 
A high thermal conduction of sapphire at low temperatures provides 
the correspondence between the thermocouple readings and sample 
temperature to a fairly high accuracy (the temperature 
difference between the sample and thermocouple 
junction evaluated below is
within 0.05 K for T-90 K).
\begin{figure}
\begin{center}
\hspace*{-6.0cm}
\parbox{8cm}{\epsfxsize=8.cm \epsfysize=7.5cm \epsfbox [5 5 500 500]
{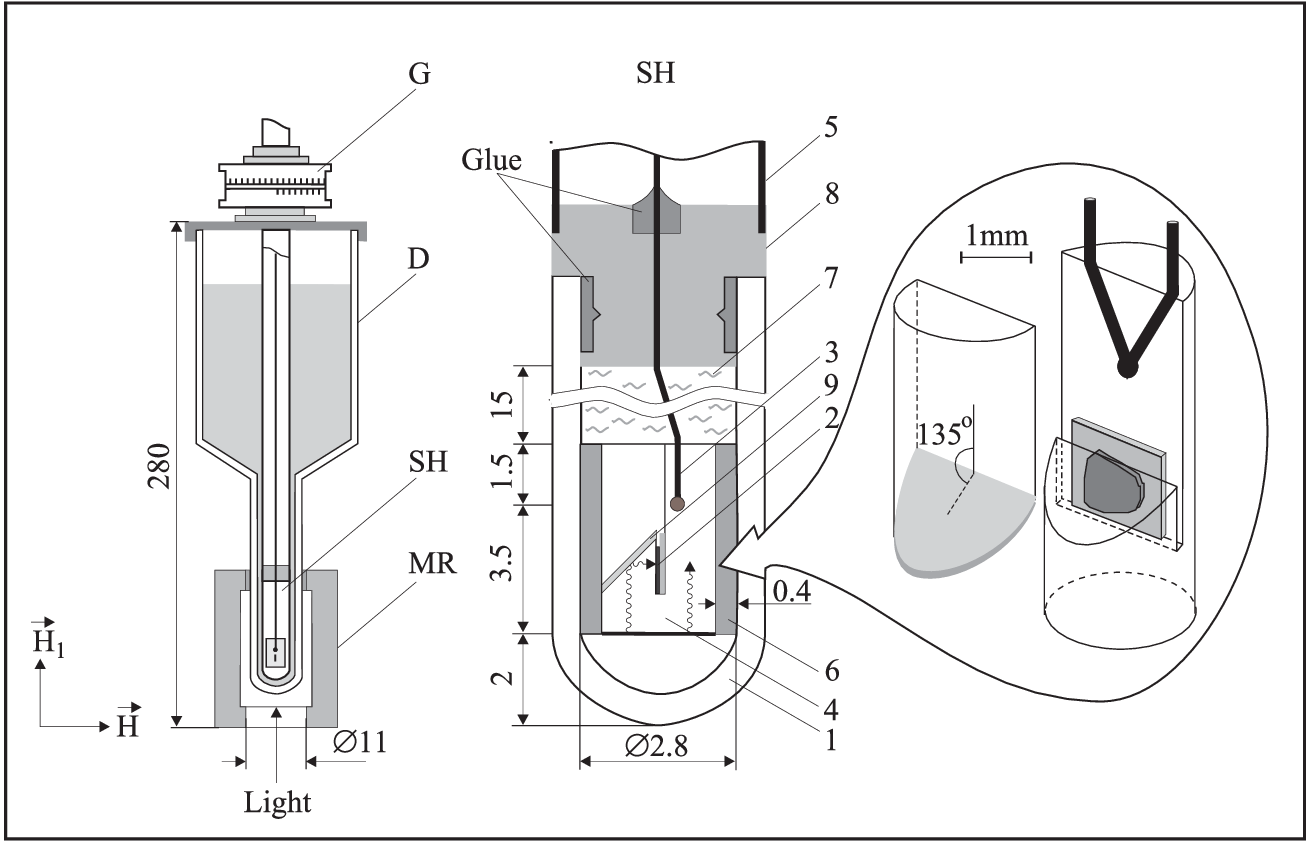}{}}
\vskip -2.0cm
\end{center}
\begin{center}
\parbox{9.5cm}{\small 
Fig.1. Designs of the cryosystem and sample holder: (1) quartz capsule; 
(2) sample; (3) thermocouple junction; (4) sapphire heat
exchanger; (5) rod of the sample holder; (6) heat insulator; (7) cotton 
layer; (8) ebonite plug; (9) reflector of the heat wave; (D) dewar;
(SH) sample holder; (MR) microwave resonator; and (G) goniometer.}
\end{center}
\end{figure}
A sample is glued to a backing of a dense paper with
the thickness that provides mechanical pressing of the
sample plane against the heat exchanger (against the
left wall of the saw cut in Fig. 1). To create a reliable
thermal contact with the heat exchanger, a thin silicone
grease layer is deposited on the sample surface.

The sample plane is oriented along the direction of
the magnetic component {$\bf H_1 $}, of the microwave field.
Such an orientation of the sample is optimum for the
study of anisotropic effects, because this allows us to
vary the angle between the sample plane and external
magnetic field H within the maximal possible range by
rotating the capsule about its axis. Rod 5 of the sample
holder (n thin-walled tube made of stainless steel) is
fixed by a collet type clip of the goniometric head (G),
thus making it possible to turn the capsule around the
vertical axis through an arbitrary angle.

The sample is heated to a temperature above the
nitrogen one by the light flux, which is focused on the
blackened bottom end of the heat exchanger by using a
system of lenses. A KGM-150 halogen lamp serves as
a light source. Its power and, consequently, the light
flux intensity incident on the heat exchanger is controlled 
with the help of a regulated dc power supply.
The minimal step in temperature variations is within
0.06 K.

The efficiency of sample heating depends on the
heat insulator 6 material used, which plays the principal
role in the heat exchange between the sapphire cylinder
and thermostat (liquid nitrogen $ N_2 $ ). The quality of thermal
contact can be varied by changing the heat insulator material.
A layer of fluffed up cotton 7 filling the
space between the heat exchanger and ebonite plug 8 of
capsule 1 prevents from the appearance of thermal convective 
air flows inside the capsule. When cotton was
used as a heat insulator 6, the sample temperature
reached a value of 180 K at the maximum lamp power.

The stability of the sample temperature at a given
current through the lamp filament is strongly dependent
on the capsule sealing and liquid nitrogen purity. As
was experimentally noted, the instability caused by liquid 
nitrogen overheating is efficiently removed by a
well-known method of using a cotton thread immersed
into liquid nitrogen down to the level of the bottom end
of capsule l. If the appropriate conditions were satisfied, 
the rms instability of the sample temperature was
within 0.06 K over a 5-min interval.

The sample temperature modulation is implemented
by 80-Hz chopping of the light flux with a two-lobe
obturator, which also controls the operation of an optical 
sensor (photodiode-light-emitting diode) generating 
the reference signal for the synchrodetector of the
receiving and measuring circuit. The choice of this frequency 
is connected with an appropriately modernized
80 Hz UNIT block, which is included in the ESR spectrometer 
set and normally used for low-frequency magnetic modulation 
with a frequency of 80 Hz. Here, this
block serves as a synchrodetector.

An IBM PC/AT computer connected to the
CAMAC bus controls the system operation, monitors
its parameters, and performs the acquisition, processing, 
and visualization of the data obtained.

Owing to the simplicity of the thermoregulating system, 
its most important characteristics can be determined
by computations. Below we give the appropriate estimates 
obtained by using the reference data from \cite{ref7} - \cite{ref8}.

To evaluate the heat flux $\win(\tso)$ conveyed to the
bottom end of the sapphire cylinder for maintaining its
average temperature $\tso$, we assume that because of a
high thermal conductivity of sapphire ($\lambda\approx700\tpr $)
for $T\approx90K$), the temperature at an arbitrary point of
the cylinder is close to its average value. Then, in view
of the fact that under the steady-state conditions
$\win(\tso)$ is equal to the heat flux $\wou(\tso)$ leaving the
cylinder through its surface, we can write
\begin{equation}
\label{1}
\win(\tso)=\wou(\tso)\approx\frac{\displaystyle\lambda'S}
{\displaystyle \delta}(\tso-T_a),
\end{equation}
where $\lambda'$ is the thermal conductivity of the heat insulator 
in the gap between the lateral surface of the cylinder
and the capsule wall, $\delta$ is the gap value, S is the area of
the lateral surface of the cylinder, and $T_a$ is the temperature 
of the internal surface of the capsule that, according to 
crude estimates, can be assumed equal to the
temperature of liquid nitrogen. On the right-hand side
of the latter equality, we omitted the terms that correspond 
to the heat fluxes leaving the cylinder through its
upper end and through the thermocouple wires,
because their overestimated total contribution is no
larger than 10\%. The factor $\alpha\equiv{\lambda'S}/{\delta}$ 
for the heat insulator 
in the form of paper ($\lambda'\approx 4 \cdot 10^{-2}\tpr $) and
air $\alpha\approx 3 \cdot 10^{-3}{W}/{K} $ 
takes on the value ($\lambda'\approx 10^{-2}\tpr $)
and $\alpha\approx 8 \cdot 10^{-4}{W}/{K} $, respectively.

Now it is easy to assess the uniformity of the temperature 
distribution over the sapphire cylinder volume.
Taking into account that the heat flux density and, consequently, 
the temperature gradient $ \gr $ inside the cylinder volume is 
maximum near the bottom cylinder end, we can write the inequality
\begin{equation}
\label{2}
\gr\le\frac{\displaystyle\win(\tso)}{\displaystyle\lambda\sigma}
\approx\frac{\displaystyle\alpha}{\displaystyle\lambda\sigma}(\tso-T_a),
\end{equation}
where $\sigma$ is the area of the cylinder end and $\lambda$ is the thermal 
conductivity of sapphire. In contrast to the heat
insulator, the thermal conductivity of sapphire strongly
depends on the temperature in the region we are interested in 
and changes from $ \lambda\sim 10^3 \tpr $ for $ T\sim 80 K $
to $ \lambda\sim 10^2 \tpr $ for $ T\sim 120 K $. From ((\ref{2})) it follows
that for a paper heat insulator, $\gr\le 2\cdot10^{-2}\tgr$ for
$\tso\approx 90 K$ and $\le 5\cdot10^{-1}\tgr$ 
for $\tso\approx 120 K$, and in case
the air is used for heat insulation,
$\gr \le 6\cdot10^{-3}\tgr$ for $\tso\approx 90 K$ and
$\gr\le 10^{-1}\tgr$ for $\tso\approx 120 K$.
We note for comparison that in the best flow-type cryostats used
as temperature attachments to ESR spectrometers, the
temperature gradient at the location of a sample can
reach a value of 0.5 K/mm within the temperature range
examined \cite{ref6}. The temperature difference in a sample
with linear dimensions of $\sim$ 1 mm that corresponds to
this value turns out to be comparable with the width of
the superconducting transition $\Delta T \sim 0,5-1K$, which is
characteristic of HTSC 1-2-3 single crystals with the
critical temperature $T_c \sim 90K$. It is obvious that the
shape of the superconducting transition peak in this situation 
must be strongly distorted in comparison with
the true one.

The time of establishing a steady-state temperature
field in the capsule at a fixed current through the lamp
filament is governed mainly by the time of transition to
a steady state in the sapphire-thermostat system. This
time can be expressed in the known form \cite{ref9}
$$ t_r=\frac{cm}{\alpha}, $$
where c is the specific heat for sapphire and m=60 mg
is the cylinder mass. For T=90 K ($c\approx100{J}/{kg\cdot K}$),
we have $t_r\approx 2 c $ for paper as a heat insulator and 
$t_r\approx 8 c $
for the air as a heat insulator. The measurements performed 
have shown the correctness of these estimates.
Note that in flow-type cryostats, the time of establishing 
a preset temperature of a sample is a few minutes.

When a light beam is chopped with the frequency $\nu$,
the time dependence of the heat flux $\win(t)$ transferred
to the cylinder has a meander form, so that the following 
expansion is valid:
$$\win(t)=W_0+\frac{\displaystyle4W_0}{\displaystyle\pi}
(\cos \omega t-\frac{\displaystyle1}
{\displaystyle 3}\cos 3 \omega t +\cdots),$$
where $\omega=2\pi\nu.$ The constant component $W_0$ of this
expansion has the physical meaning identical with that
of $\win(\tso)$ in (\ref{1}), i.e., $W_0=\win(\tso)$. 
In this case, $\tso$
denotes the cylinder temperature averaged not only
over the volume, but over time as well. The amplitude
of the fraction of the flux that oscillates with its fundamental 
frequency can be found in the form $W_\omega = {4}/{\pi}\win(\tso)$.
This allows us to evaluate the amplitude
of the temperature wave 
\linebreak $\tau_0 exp \left[i(kx-\omega t)\right]$ generated in
the direction of the cylinder axis by using the known
formula \cite{ref10}
$$\tau_0 = - \frac{\displaystyle W_\omega}
{\displaystyle\sigma\lambda\cdot ik}=
\frac{\displaystyle4}{\displaystyle\pi}
\frac{\displaystyle\win(\tso)\cdot i}
{\displaystyle\sigma\lambda k},$$
where $k=(1+i){({c\rho\omega}/{2\lambda})}^{1/2}$ is the
complex wavenumber ($\rho$ is the sapphire density). For
$\tso\sim 90K$, we have 
$\left|\tau_0 \right| \approx 4\cdot10^{-2}K$
for a paper heat insulator and $\left|\tau_0 \right| \approx 10^{-2}K.$
for the air heat insulator.
It should be noted that as a
result of damping, at the level of the sample location the
wave is expected to have smaller amplitudes than indicated above,
because the penetration depth
for sapphire is $l={({\bf\rm Im} \phantom{'} k)}^{-1}\approx 2mm $,
which slightly
exceeds the distance from the bottom end of the cylinder 
to the sample. Moreover, it was observed experimentally that 
the temperature modulation amplitude in
thin plates of HTSC single crystals with the width
\linebreak $d<0,05mm$ is significantly dependent on the angle
between the plate plane and the direction of wave propagation 
in sapphire, i.e., the angle of slide: the height of
the peak of the superconducting transition, from which
the temperature modulation amplitude was evaluated,
was at a noise level for zero angles and rapidly
increased with the angle increase. A small modulation
amplitude at zero angles of slide means that in these
cases the amplitude of the temperature wave entering
the sample is suppressed. As a consequence, we infer
that the orientation of the sample plane along the cylinder 
axis used in experiments is less favorable from the
viewpoint of creating a temperature modulation and
conserving the initial direction of the temperature
wave. This difficulty is eliminated by reflector 9
(Fig. 1), Its role is played by the boundary between sapphire 
and cardboard spacer, the thermal conductivities
of which have a ratio of $\sim10^3$. The cardboard spacer has
a much larger thickness ($d\approx 0,1mm$) than the penetration 
depth (for cardboard, l=0.02 mm) and guarantees
that the reflectance is close to unity. The reflector plane
makes an angle of $45^\circ$ with the cylinder axis and provides 
not only the normal incidence of the reflected
wave on the sample surface, but also a high uniformity
of the amplitude and phase of the temperature modulation 
along this plane, because all the imaginary rays
reaching the sample after being reflected have identical
path lengths (a wavy line with an arrow in Fig. 1 shows
the direction of the temperature wave propagation). The
temperature modulation uniformity over the sample
thickness is achieved by using another reflector, represented 
by the boundary with a paper spacer that presses
a sample against sapphire.

It can be demonstrated that the reflected wave,
which originates due to a strong decrease in the heat
conduction at the sample-paper interface, interferes
with the primary wave and must effectively smooth the
amplitude and phase of temperature oscillations inside
the sample. This statement is valid only for the samples,
the thickness of which is small enough in comparison
with the penetration depth (as a rule, HTSC single crystals 
with the thickness $d < 0.05 mm$ satisfy this requirement). 
Because of a large decrease in the heat conduction at the 
sapphire-sample interface, the amplitude of
the resulting wave inside the sample must exceed that
of the wave incident onto this interface by a factor of
almost 2. This gain satisfactorily compensates for the
temperature wave damping on the way to the sample.
Therefore, the above-mentioned values of 
$\left|\tau_0 \right|$ can be
used as estimates for the temperature modulation
amplitude in the sample.

\section{Results}
\indent
  
To illustrate the functional feasibilities of the cryosystem 
described, Fig. 2 presents the results of measuring the first 
derivative of the microwave power R
absorbed by an $YBa_2Cu_3O_{7-x}$ single crystal measuring
${\rm\sim1,0 x 0,7 x 0,03 mm}$. 
\begin{figure}
\vskip -4cm
\begin{center}
\hspace*{-5.0cm}
\parbox{8cm}{\epsfxsize=8.cm \epsfysize=6.cm \epsfbox [5 5 500 500]
{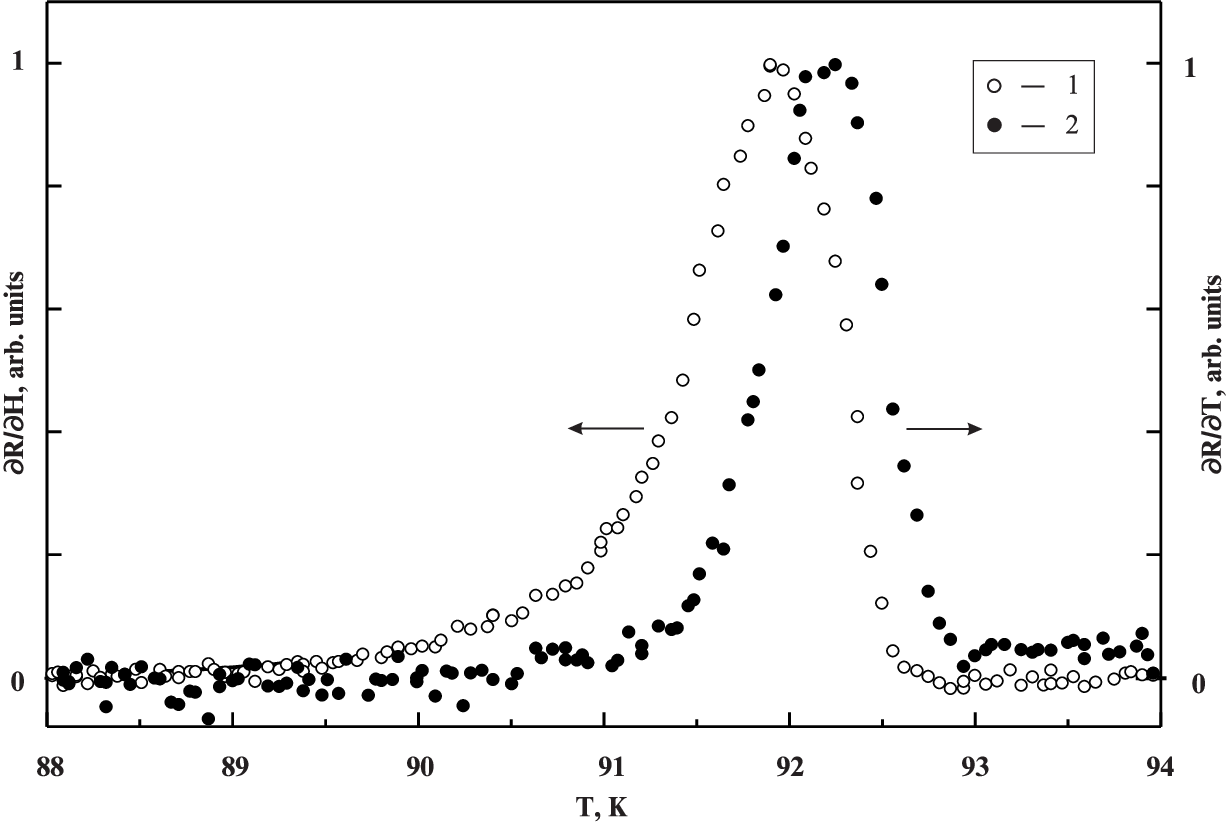}{}}
\vskip -1.5cm
\end{center}
\begin{center}
\parbox{8.5cm}{\small 
Fig.2. Temperature-dependent microwave absorption in an
$YBa_2Cu_3O_{7-x}$, single crystal measured by the (1) magnetic
and (2) temperature modulation methods.}
\end{center}
\end{figure}
The results were obtained according 
to the conventional technique with the modulation
of an external magnetic field strength (the frequency is
100 kHz and modulation amplitude is 10 Oe) and with
our method using the sample temperature modulation
(the frequency is 80 Hz and modulation amplitude is
$10^{-2} \div 10^{-1}K$ ). Measurements were performed under the
conditions of cooling the sample from above-critical
temperatures in the external field H = 20 Oe directed
perpendicular to the sample plane.

In summary it should be noted that the initial, less
perfect version of this system of thermal regulation was
applied to the investigations of microwave absorption
in HTSC ceramics \cite{ref2}.

\end{document}